**Title**

# A 3D deep learning classifier and its explainability when assessing coronary artery disease

Wing Keung Cheung[1], Jeremy Kalindjian[2], Robert Bell[2], Arjun Nair[3], Leon J. Menezes[4], Riyaz Patel[5], Simon Wan[4], Kacy Chou[1], Jiahang Chen[6], Ryo Torii[6], Rhodri H. Davies[5,7], James C. Moon[7], Daniel C. Alexander[1], Joseph Jacob[1,8]

[1]Satsuma Lab, Centre for Medical Image Computing & Department of Computer Science, University College London, London, UK

[2]Hatter Cardiovascular Institute, University College London, London, UK

[3]Department of Radiology, University College London Hospital, London, UK

[4]Institute of Nuclear Medicine, University College London, London, UK

[5]Institute of Cardiovascular Science, Faculty of Population Health Sciences, University College London, London, UK

[6]Department of Mechanical Engineering, University College London, London, UK

[7]Barts Heart Centre, West Smithfield, London, UK

[8]Department of Respiratory Medicine, University College London, London, UK

Corresponding author:

Dr Joseph Jacob

UCL Centre for Medical Image Computing

1st Floor, 90 High Holborn, London WC1V6LJ

j.jacob@ucl.ac.uk

## Abstract

Early detection and diagnosis of coronary artery disease (CAD) could save lives and reduce healthcare costs. The current clinical practice is to perform CAD diagnosis through analysing medical images from computed tomography coronary angiography (CTCA). Most current approaches utilise deep learning methods but require centerline extraction and multi-planar reconstruction. These indirect methods are not designed in a clinician-friendly manner, and they complicate the interventional procedure. Furthermore, the current deep learning methods do not provide exact explainability and limit the usefulness of these methods to be deployed in clinical settings. In this study, we first propose a 3D Resnet-50 deep learning model to directly classify normal subjects and CAD patients on CTCA images, then we demonstrate a 2D modified U-Net model can be subsequently employed to segment the coronary arteries. Our proposed approach outperforms the state-of-the-art models by 21.43% in terms of classification accuracy. The classification model with focal loss provides a better and more focused heat map, and the segmentation model provides better explainability than the classification-only model. The proposed holistic approach not only provides a simpler and clinician-friendly solution but also good classification accuracy and exact explainability for CAD diagnosis.

## Introduction

Coronary artery disease (CAD) is a common cause of death [1] in developed (i.e., UK and USA) and developing countries (i.e., India). Early detection and diagnosis of CAD could save lives and costs [2]. Currently, computed tomography coronary angiography (CTCA) plays a central role in diagnosing or excluding CAD in patients with chest pain [3, 4]. It has the ability to image stable and unstable atherosclerotic plaques which can result in coronary artery stenosis. The assessment of CAD severity is based on a clinicians' visual assessment of the severity of coronary artery narrowing on automated segmentation. This procedure, however, is subjective [5, 6] and the accuracy of the assessment is influenced by the experience of the clinician. Given the large number of suspected CAD patients in the NHS and the existing shortage of clinicians available to review the CTCA [7], it is crucial to develop fast, efficient, objective and accurate automated detection and classification systems to assist clinicians to diagnose CAD. This will also enable faster triage of patients using computer technology on local hospital servers, speeding up diagnosis and patient management.

Deep learning [8] is a promising method to deliver fully automatic diagnosis for CAD. It utilises image data and GPU technology to provide fast and efficient computation. Deep learning based CAD diagnosis on CTCA can be broadly divided into two categories, (1) CAD segmentation [9] (2) CAD classification [10]. CAD segmentation aims to produce a segmented mask containing the proximal ascending aorta and the coronary arteries. The mask is particularly useful for visual estimation of stenosis severity. Furthermore, it can be used for computational fluid dynamics (CFD) [11] which can estimate the blood flow in the coronary vessels. This clinical information is important for diagnosing hemodynamically significant CAD. On the other hand, CAD classification aims to grade CAD severity (i.e., no CAD, non-obstructive CAD and obstructive (severe) CAD) [12]. This can be achieved by predicting a severity label at the patient-level. A deep learning based classifier may use CTCA images and learn the non-linear mapping between image features and CAD severity grades. It offers robust and efficient classification and can be implemented easily on local hospital severs.

Deep learning methods have been implemented successfully for medical image segmentation and classification. However, one of the drawbacks of deep learning is that its black box nature limits its interpretability and therefore clinical applicability as end users may struggle to understand the decisions of the classifier. Therefore, the deep learning research community has increasingly focused on explanability. Explainable AI (XAI) [13] aims to provide explainable tools for deep learning methods. This can improve confidence and trust for end users in deep learning output. In the context of CTCA image classification, a popular XAI tools is Grad-GAM [14]. It allows direct visualisation of regions in the image that the deep learning model found to be important. The Grad-CAM employs gradient-weighted class activation mapping and provides visual explanations for deep learning based classifiers.

Currently, no studies focus on direct (without centreline extraction and multi-planar reconstruction (MPR)) and explainable approaches for CAD classification. In this study, we propose a 3D deep learning approach with explainability for the automated classification of normal subjects and CAD patients on CTCA images.

The main contributions of this study are:

- The first study to propose a 3D Resnet-50 deep learning model that directly (without centreline extraction and MPR) classifies normal subjects and patients with coronary artery disease on CTCA images with good accuracy and provides good explainability that aligns with clinicians' diagnostic criteria.

- Compared with 3D DenseNet-121 and 3D EfficientNetB0 models, our proposed method outperformed the state-of-the-art models by 21.43%.

- Our proposed method is fast, efficient and particularly good at classifying normal subjects on CTCA images and requires only subject-level labels.

- In terms of classification explainability, the study confirms that focal loss is better than binary cross entropy, and the model with focal loss focused on the heart including coronary arteries.

- Our study demonstrates the 3D CAD classification can be further formulated as 2D two-class semantic segmentation, and it provides exact explainability and accurate segmented mask of coronary arteries.

The organisation of the rest of the paper is structured as follows: Section II reviews the latest and relevant research work regarding CAD classification. The methods of the proposed work are presented in Section III. The experimental results are shown in Section IV. Section V discusses the research findings and addresses the potential implications, limitations, and future research directions. Finally, Section VI summarises and concludes the key findings, contribution, and potential impact of the proposed work.

## Related work

Previous studies related to deep learning based CAD segmentation are summarised in [15]. More recently, Lin et al. [16] have proposed the hierarchical convolutional long short-term memory (ConvLSTM) network to segment the coronary arteries. It demonstrated excellent agreement between deep learning and expert readers for volumes of total plaque, calcified plaque and non-calcified plaque. Song et al. [17] have developed a three-stage approach to segment the coronary arteries. A 2D DenseNet was first employed to classify coronary arteries and non-coronary arteries. Then, a 3D FFR-Unet was used to segment the initial coronary arteries. The final segmentation results were obtained by performing Gaussian weighted averaging on the segmentation predictions.

The studies related to deep learning based CAD classification are summarised in this review [10]. Furthermore, Raghav et al. [18] demonstrated an automated CNN model based on a Shufflenet architecture for detecting CAD from CTCA with high accuracy. The drawback of this approach is that it requires slice-level annotations, which are time consuming and labour intensive. Furthermore, the model is not able to localise the abnormality in the coronary artery. Zreik et al. [19] proposed a 3D recurrent CNN for automatic detection and classification of coronary artery plaque and stenosis in coronary CT angiography. The drawback of this method is that it requires extraction of a centerline and the 3D CNN model requires high GPU memory. Candemir et al. [20] developed a 3D CNN deep learning model to classify the coronary arteries and localise abnormal regions. It also requires centerline extraction and high GPU memory. The above approaches reduce the computational and GPU memory efficiency.

The studies related to explainability of deep learning based CAD classification are limited. Salih et al. [21] reviewed XAI methods in cardiac imaging as well as explainable AI with applications in cardiology [22]. Though the study considered cardiac magnetic resonance (CMR) imaging, the XAI techniques are applicable to CTCA. Candemir et al. [20] used a weakly supervised method to localise coronary abnormalities, thereby confirming that the 3D CNN model focused on stable and unstable plaques. However, abnormality localisation was still not exact.

## Methods

### *Patient data*

The data were collected from University College Hospital London and Barts Health NHS Trust. CTCA scans were performed on 88 subjects. All the patients in the study had been evaluated for possible angina, with an intermediate risk of coronary artery disease. All patients underwent cardiac CT angiography for anatomical assessment of their coronary arteries and risk stratification for coronary artery disease. The study was carried out in accordance with the recommendations of the South East Research Ethics Committee, Aylesford, Kent, UK, and Leeds East Research Ethics Committee: 20/YH/0120 with written informed consent from all subjects in accordance with the Declaration of Helsinki.

### *Data pre-processing*

The CTCA images were pre-processed using ImageJ (1.53). The intensity of the images were normalised using linear histogram stretch and subsequently rescaled to between 0 to 255. The images were resized from 512x512 to 128x128. The processed images were saved to 8-bit Portable Network Graphics (PNG). The images were used to feed the deep learning models for training, validation and testing. It should be noted that the total number of slices (per case) was resampled to 256 for 3D CAD classification.

### *Data annotation*

An experienced cardiologist performed the data annotation. The labels were produced at the subject level. The CAD classification is based on degree of coronary artery stenosis: 0 = normal, 1 = mild stenosis (<50%), 2 = moderate stenosis (50%–70%), and 3 = severe stenosis (>70%). When abnormal features (stenosis with/without calcification) were observed in subjects' CT volumetric images, CAD (abnormal) labels were assigned to the subjects' CT images. Otherwise, normal labels were assigned.

### *2D CAD classification*

The 2D Resnet-50 was used to perform 2D CAD classification. It is a 50-layer convolutional network and utilises shortcut connections to form a residual network. Shortcut connections improve the information and gradients to flow more easily throughout the network. The network adopts a bottleneck design strategy, which has 1x1 convolutions. This reduces the number of parameters and matrix multiplications and therefore allows faster training on each layer. Figure 1 shows the network architecture of 2D Resnet-50.

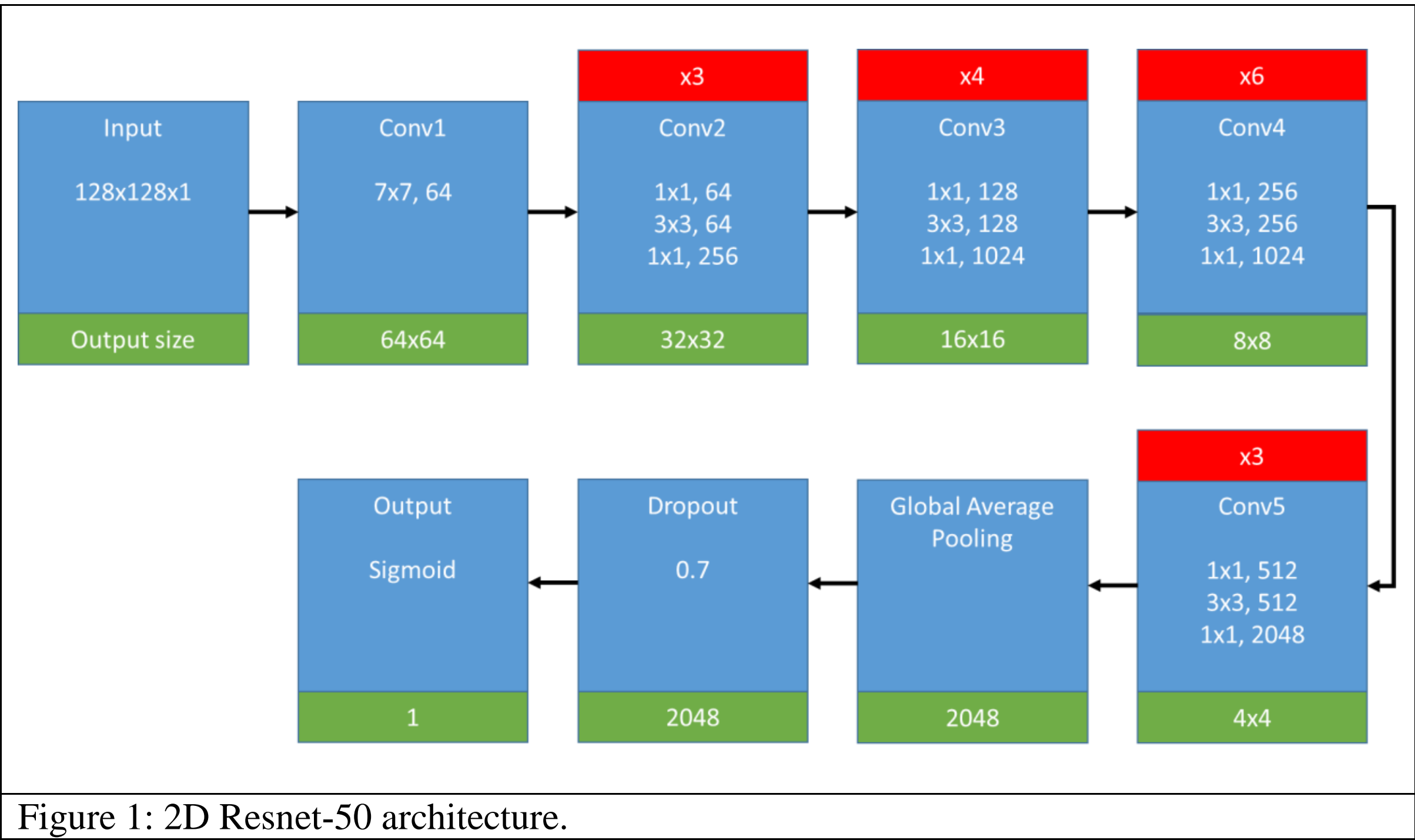

Figure 1: 2D Resnet-50 architecture.

*3D CAD classification*

3D CAD classification was performed using 3D Resnet-50. The 3D Resnet-50 network is the extension of the 2D Resnet-50 network. It uses 1x1x1 convolutions for bottleneck block. It has the ability to learn features across several images while the 2D Resnet-50 learns the features within a single image only. The network architecture of 3D Resnet-50 is shown in Figure 2.

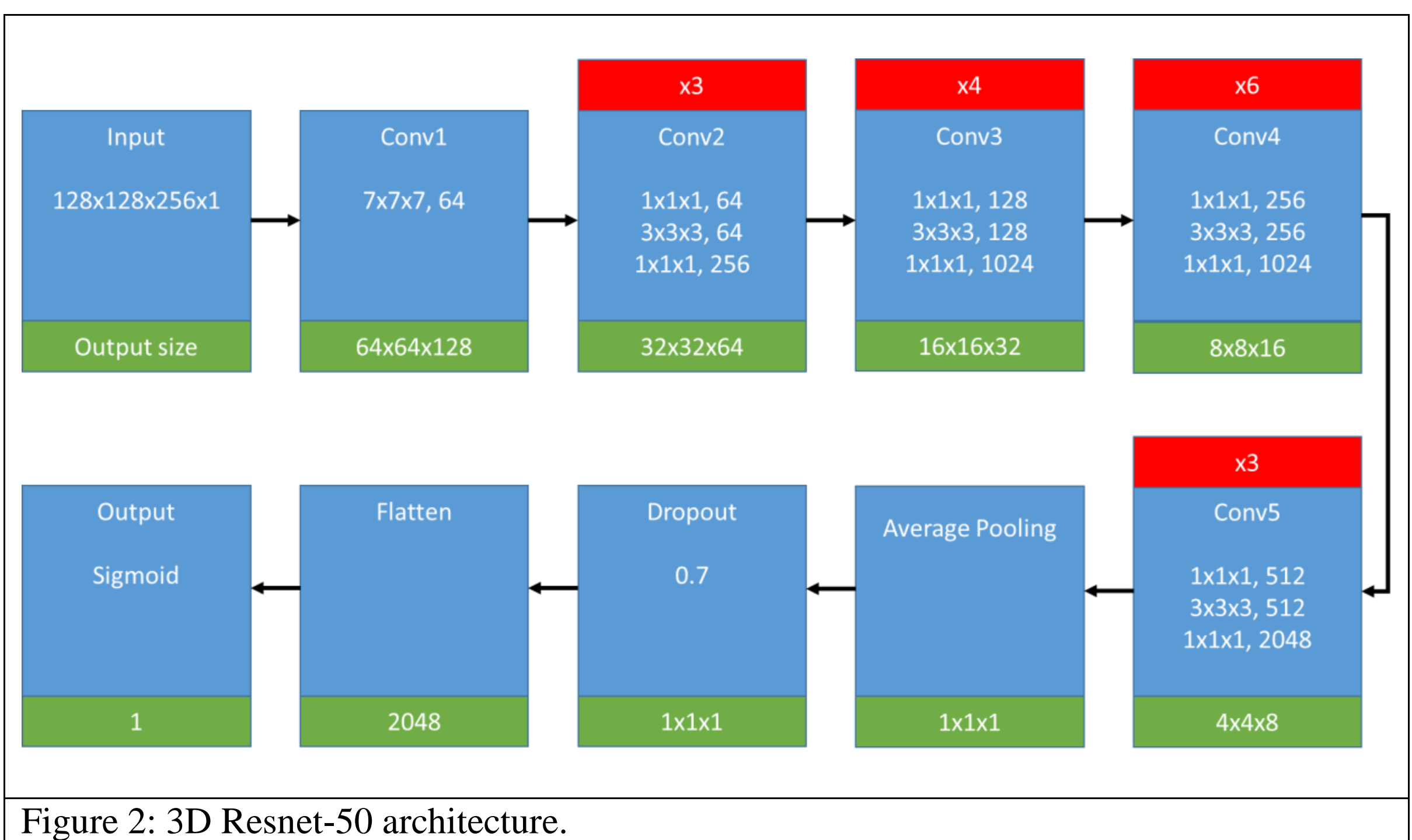

Figure 2: 3D Resnet-50 architecture.

*Model training and implementation*

The clinical data was split into training, validation and test sets. The ratio of normal subjects to CAD patients was maintained for each set. Table 1 shows the number of normal subjects and CAD patients in each set. Five-fold cross-validation was implemented for parameters selection. The final models were selected and evaluated on the test set. Moreover, the corresponding total slices for each set is summarised in Table 2. It should be noted that 3D classification is performed at a subject level, while 2D classification is performed at a CT slice level.

Table 1: Cases allocation (per fold) for training, validation and test sets.

| **Cases (per fold)** | **Normal subject** | **CAD patient** | **Total** |
|---|---|---|---|
| Training set | 30 | 30 | 60 |
| Validation set | 7 | 7 | 14 |
| Test set | 7 | 7 | 14 |
| *Total* | 44 | 44 | 88 |

Table 2: Slices allocation (per fold) for training, validation and test sets.

| **Number of slices** | **Normal subject** | **CAD patient** | **Total** |
|---|---|---|---|
| **Fold 1** | | | |
| Training set | 8173 | 7762 | 15935 |
| Validation set | 2200 | 2014 | 4214 |
| Test set | 1946 | 2153 | 4099 |
| *Total* | 12319 | 11929 | 24248 |
| | | | |
| **Fold 2** | | | |
| Training set | 8363 | 7952 | 16315 |
| Validation set | 2010 | 1824 | 3834 |
| Test set | 1946 | 2153 | 4099 |
| *Total* | 12319 | 11929 | 24248 |
| | | | |
| **Fold 3** | | | |
| Training set | 8482 | 7926 | 16408 |
| Validation set | 1891 | 1850 | 3741 |
| Test set | 1946 | 2153 | 4099 |
| *Total* | 12319 | 11929 | 24248 |
| | | | |
| **Fold 4** | | | |
| Training set | 8532 | 7967 | 16499 |
| Validation set | 1841 | 1809 | 3650 |
| Test set | 1946 | 2153 | 4099 |
| *Total* | 12319 | 11929 | 24248 |
| | | | |
| **Fold 5** | | | |
| Training set | 8608 | 8047 | 16655 |
| Validation set | 1765 | 1729 | 3494 |
| Test set | 1946 | 2153 | 4099 |
| *Total* | 12319 | 11929 | 24248 |

The proposed Resnet-50 models were implemented in Tensorflow (2.15.0). They were executed on four high performance machines with the following configurations: (1) 2x16-Core Intel Xeon E5-2683V3 @2.00 GHz with a NVIDIA Tesla V100, (2) 2x16-Core Processor Intel Xeon Gold 6326 @2.90 GHz with a NVIDIA Hopper H100, (3) 2x24-Core Intel Xeon Gold 6326 @2.90 GHz with a NVIDIA Tesla A100, and (4) 2x32-Core Intel Xeon Platinum 8462Y+ @2.80 GHz with a NVIDIA Hopper H100. The optimal model was chosen with the maximum validation accuracy. The parameter settings are shown in Table 3.

Table 3: The parameter settings of 2D and 3D Resnet-50 models

| | **2D Resnet-50** | **3D Resnet-50** |
|---|---|---|
| **Optimiser** | Adam | Adam |
| **Learning rate** | 0.001 | 0.001 |
| **Batch size** | 10 | 1 |
| **Epochs** | 100 | 10 |
| **Early stopping** | | |
| *Patience* | 10 | NA* |
| **ReduceLROnPlateau** | | |
| *Minimum learning rate* | 0.00001 | NA* |
| *Factor* | 0.1 | NA* |
| *Patience* | 3 | NA* |

*NA – Not applicable

*Loss functions and performance evaluation*

In this study, binary cross entropy (BCE) and focal loss (FL) were employed for training the deep learning models. BCE is commonly used for binary classification, it utilises information entropy to measure the similarity between the predicted probability and ground-truth label. FL is used for highly imbalanced dataset, and it generalises the BCE with a focusing parameter ($\gamma$) and weighting factor ($\alpha$). $\gamma$ is introduced to penalise more heavily on hard-to-classify samples relative to easy-to-classify samples, while $\alpha$ can be used to rebalance the loss function due to class imbalance. The settings are $\alpha = 0.5$ and $\gamma = 4$. Furthermore, accuracy, recall, precision, and F1-score are used to measure the classification performance. The losses and the evaluation metrics are shown by the following mathematical equations (eqs.1-7).

$$\text{BCE loss} \;= -\frac{1}{n}\sum_{i=1}^{n}\left(\left(y_i \times log\ \hat{y}_i\right) + \left(1 - y_i\right) \times log\left(1 - \hat{y}_i\right)\right) \quad (1)$$

$$sample^{FL} = \begin{cases} -\alpha \times \left(1 - \hat{y}_i\right)^{\gamma} \times log\ \hat{y}_i\ , for\ positive\ label\ prediction \\ -\left(1 - \alpha\right) \times {\hat{y}_i}^{\gamma} \times \log\left(1 - \hat{y}_i\right), for\ negative\ label\ prediction \end{cases} \quad (2)$$

$$\text{focal loss} \;= \frac{1}{n}\sum_{i=1}^{n} sample_n^{FL} \quad (3)$$

Where $y_i$ is the ground-truth label, $\hat{y}_i$ is the predicted probability of positive label, $n$ is the total number of samples, $\alpha$ is the weighing factor, and $\gamma$ is the focusing parameter.

$$\text{accuracy} = \frac{TP+TN}{TP+TN+FP+FN} \quad (4)$$

$$\text{recall} = \frac{TP}{TP+FN} \quad (5)$$

$$\text{precision} = \frac{TP}{TP+FP} \quad (6)$$

$$\text{F1-score} = 2 \times \frac{recall \times precision}{recall + precision} \quad (7)$$

Where $TP$ is the true positive, $TN$ is the true negative, $FP$ is the false positive, and $FN$ is the false negative.

*Classification Explainability*

The explainability of the proposed methods was investigated by Grad-CAM [14]. The last convolution layers from the proposed models were selected for this study. The Grad-CAM produces a heat map and highlights the regions in the image which are important for the classification. Given a 3D classification network with output $y^c$, representing the score for class $c$, and the goal is to compute the Grad-CAM map for a convolutional layer with $k$ feature maps (channels), $A^k_{x,y,z}$, where $x$, $y$, $z$ are the indexes the voxels. The neuron importance weight is computed as:

$$a^k_c = \frac{1}{N} \sum_x \sum_y \sum_z \frac{\partial y^c}{\partial A^k_{x,y,z}} \quad (8)$$

where $N$ is the total number of voxels in the feature map.

The Grad-CAM map, $M^c$, is a weighted combination of the feature maps with an applied ReLU:

$$M^c = ReLU\left(\sum_k a^k_c A^k\right) \quad (9)$$

*Segmentation Explainability*

An alternative approach to CAD classification is semantic segmentation. 3D CAD classification was treated as a 2D two-class (aorta and coronary arteries vs. non-aorta and non-coronary arteries) semantic segmentation. The segmentation of aorta and coronary arteries and its explainability were done in our previous studies [21, 22]. A 2D modified U-Net was employed to segment the target tissues. Furthermore, the explainability of this method was investigated. Firstly, the 2D binary segmentation was converted to a 2D two-class semantic segmentation. Then, the trained model was analysed by Seg-Grad-CAM [23]. The Seg-Grad-CAM employs gradient-weighted class activation mapping and provides a method for explaining the segmentation decision. It should be noted that our intention is not about integrating the 2D semantic segmentation into 3D classification; rather, our focus is on the usefulness of segmentation explainability and investigating whether it can provide an exact explainability than classification.

*Comparative analysis of state-of-the-art (SOTA) methods*

The proposed 3D model with FL was compared with 3D DenseNet-121 [24] and 3D EfficientNetB0 [25], respectively. These two models were commonly used for image classification. The measurements of recall, precision, F1-score, and accuracy on the test set were reported. Furthermore, the settings of these models are shown in Table 4.

Table 4: The parameter settings of 3D DenseNet-121 and 3D EfficientNetB0.

| Model | Parameter |
|---|---|
| **3D DenseNet-121** | reduction = 0.5 |
| | dropout_rate = 0.7 |
| | weight_decay = 0.0 |
| | |
| **3D EfficientNetB0** | width_coefficient = 1.0 |
| | scaling coefficient = 1.0 |
| | default_resolution = 128 |
| | dropout_rate = 0.7 |

## Results

*Classification accuracy with five-fold cross-validation*

Table 5 shows the classification accuracy of 2D and 3D Resnet-50 models. The 2D Resnet-50 model with BCE achieved 65.44% ± 1.51% and 73.79% ± 7.80% for training and validation accuracy, respectively, while the 2D Resnet-50 model with FL achieved 61.10% ± 1.19% and 73.79% ± 7.80% for training and validation accuracy, respectively. Furthermore, the 3D Resnet-50 model with BCE achieved 57.10% ± 3.14% and 72.86% ± 7.82% for training and validation accuracy, respectively, while the proposed 3D Resnet-50 model with FL achieved 61.46% ± 6.47% and 65.72% ± 9.31% for training and validation accuracy, respectively.

Table 5: The classification accuracy of 2D and 3D Resnet-50 models with five-fold cross-validation.

| Classification model | Training Accuracy ± SD | Validation Accuracy ± SD |
|---|---|---|
| 2D Resnet-50 (BCE) | 65.44% ± 1.51% | 73.79% ± 7.80% |
| 2D Resnet-50 (FL) | 61.10% ± 1.19% | 73.79% ± 7.80% |
| 3D Resnet-50 (BCE) | 57.10% ± 3.14% | 72.86% ± 7.82% |
| 3D Resnet-50 (FL) | 61.46% ± 6.47% | 65.72% ± 9.31% |

*Classification accuracy for the final classification models*

The recall, precision, F1-score and accuracy of the final 2D and 3D Resnet-50 models for test set are shown in Table 6. With regard to recall, precision, F1-score, and accuracy measurements on the test set, our proposed model outperformed the 2D Resnet-50 model by 3.49% on test accuracy. Our approach with FL achieved 0.43, 1.00, and 0.60 for recall, precision, and F1-score, respectively, while the 3D Resnet-50 model with BCE achieved 0.43, 1.00, and 0.60 for recall, precision, and F1-score, respectively. On the other hand, the 2D Resnet-50 model with FL achieved 0.39, 1.00, and 0.56 for recall, precision, and F1-score, respectively, while the 2D Resnet-50 model with BCE achieved 0.39, 1.00, and 0.56 for recall, precision, and F1-score, respectively.

Table 6: The recall, precision, F1-score and accuracy of the final 2D and 3D Resnet-50 models for test set.

| Final classification model | Recall | Precision | F1-score | Accuracy |
|---|---|---|---|---|
| 2D Resnet-50 (BCE) | 0.39 | 1.00 | 0.56 | 67.94% |
| 2D Resnet-50 (FL) | 0.39 | 1.00 | 0.56 | 67.94% |
| 3D Resnet-50 (BCE) | 0.43 | 1.00 | 0.60 | 71.43% |
| 3D Resnet-50 (FL) | 0.43 | 1.00 | 0.60 | 71.43% |

*Confusion matrix for test set*

Tables 7 and 8 show the TP, FP, FN, and TN measurements for the final 2D and 3D Resnet-50 models. On the test set, the 2D Resnet-50 model with BCE produced 20.47%, 0.00%, 32.06%, and 47.47% for TP, FP, FN, and TN, respectively, and the 2D Resnet-50 model with FL produced the same result. Our proposed model with BCE produced 21.43%, 0.00%, 28.57%, and 50.00% for TP, FP, FN, and TN, respectively. Interestingly, the proposed model with FL also produced the same result. Our proposed model (with BCE or FL) performed better than the 2D Resnet-50 model (with BCE or FL) by 0.96% and 2.53% for TP and TN, respectively, on the test set. However, in terms of FN, our proposed model performed worse than the 2D Resnet-50 model by 3.49%. Furthermore, there is no difference in FP for the both models.

Table 7: The TP, FP, FN and TN measurements for the final 2D Resnet-50 model.

| Test set (4099 slices) | TP | FP | FN | TN |
|---|---|---|---|---|
| **BCE** | 839 (20.47%) | 0 (0.00%) | 1314 (32.06%) | 1946 (47.47%) |
| **FL** | 839 (20.47%) | 0 (0.00%) | 1314 (32.06%) | 1946 (47.47%) |

Table 8: The TP, FP, FN and TN measurements for the final 3D Resnet-50 model.

| Test set (14 cases) | TP | FP | FN | TN |
|---|---|---|---|---|
| **BCE** | 3 (21.43%) | 0 (0.00%) | 4 (28.57%) | 7 (50.00%) |
| **FL** | 3 (21.43%) | 0 (0.00%) | 4 (28.57%) | 7 (50.00%) |

*Explainability*

One normal subject and one CAD patient with a correct prediction were selected to illustrate the explainability. The corresponding normal and abnormal slices were chosen accordingly. Figure 3 shows the Grad-GAM generated by the final 2D Resnet-50 model from last convolution layers on a normal subject and a CAD patient with correct prediction, respectively, while Figure 4 shows the Grad-GAM generated by the final 2D Resnet-50 model from last convolution layers on two CAD patients with incorrect prediction. Furthermore, Figure 5 shows the Grad-GAM generated by the final 3D Resnet-50 model from the last layer on a normal subject and a CAD patient with correct prediction, respectively, while Figure 6 shows the Grad-GAM generated by the final 3D Resnet-50 model from the last convolution layers on two CAD patients with incorrect prediction.

(1) 2D CAD classification by the final 2D Resnet-50 model

Regarding the correct prediction, the 2D model with BCE/FL focused on the corners of the CT image for the normal subject. On the other hand, for the CAD patient 1, the model with FL focused on the upper and left regions close to the edges, and it did not focus on the heart. Furthermore, the model with BCE focused on the heart and the lung regions. The region of calcification was captured by the model. Regarding the incorrect prediction, the model with BCE/FL did not focus on the heart and capture important features for both patients.

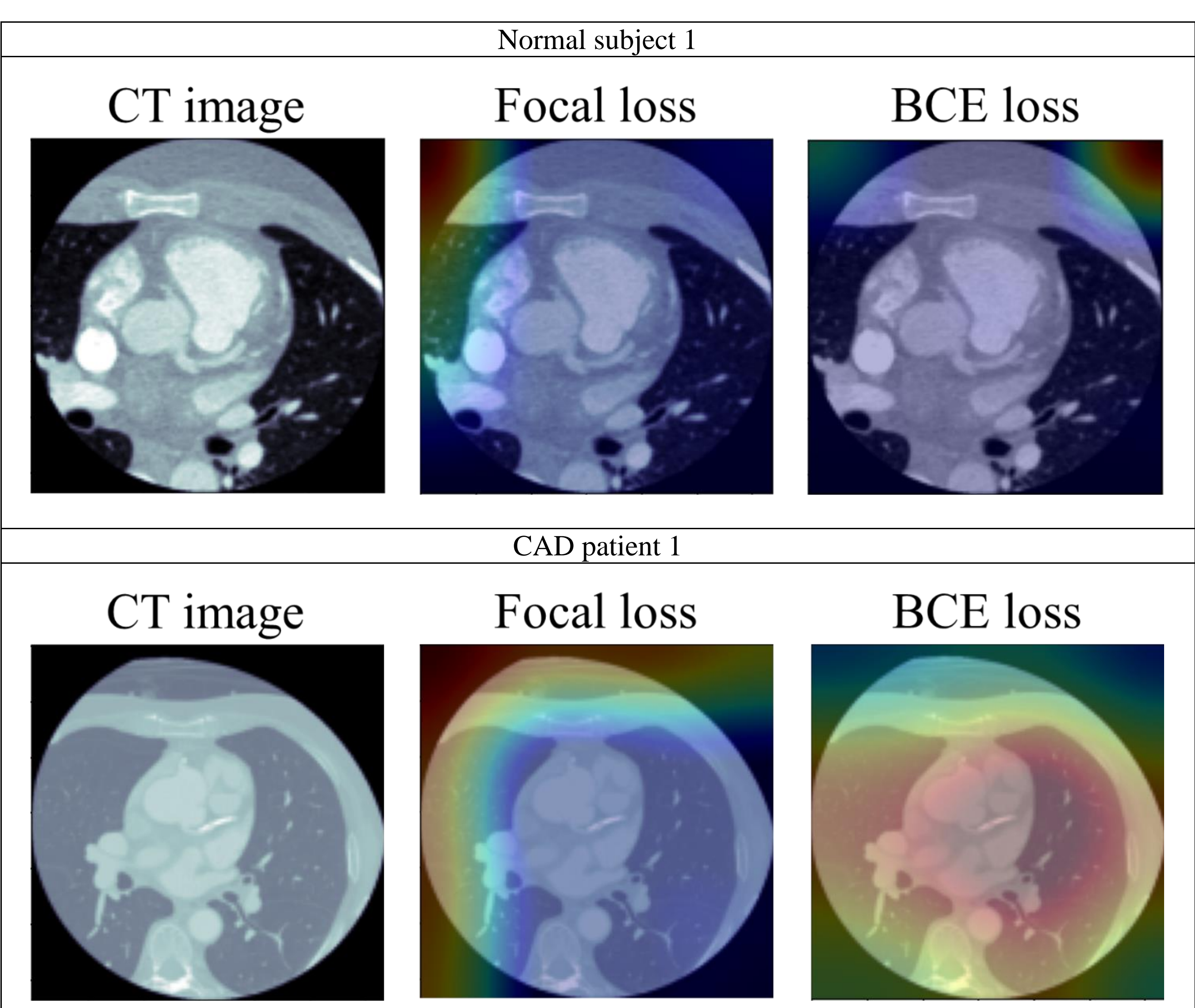

Figure 3: The Grad-GAM generated by the final 2D Resnet-50 model from last convolution layer on a normal subject 1 and a CAD patient 1 with a correct prediction.

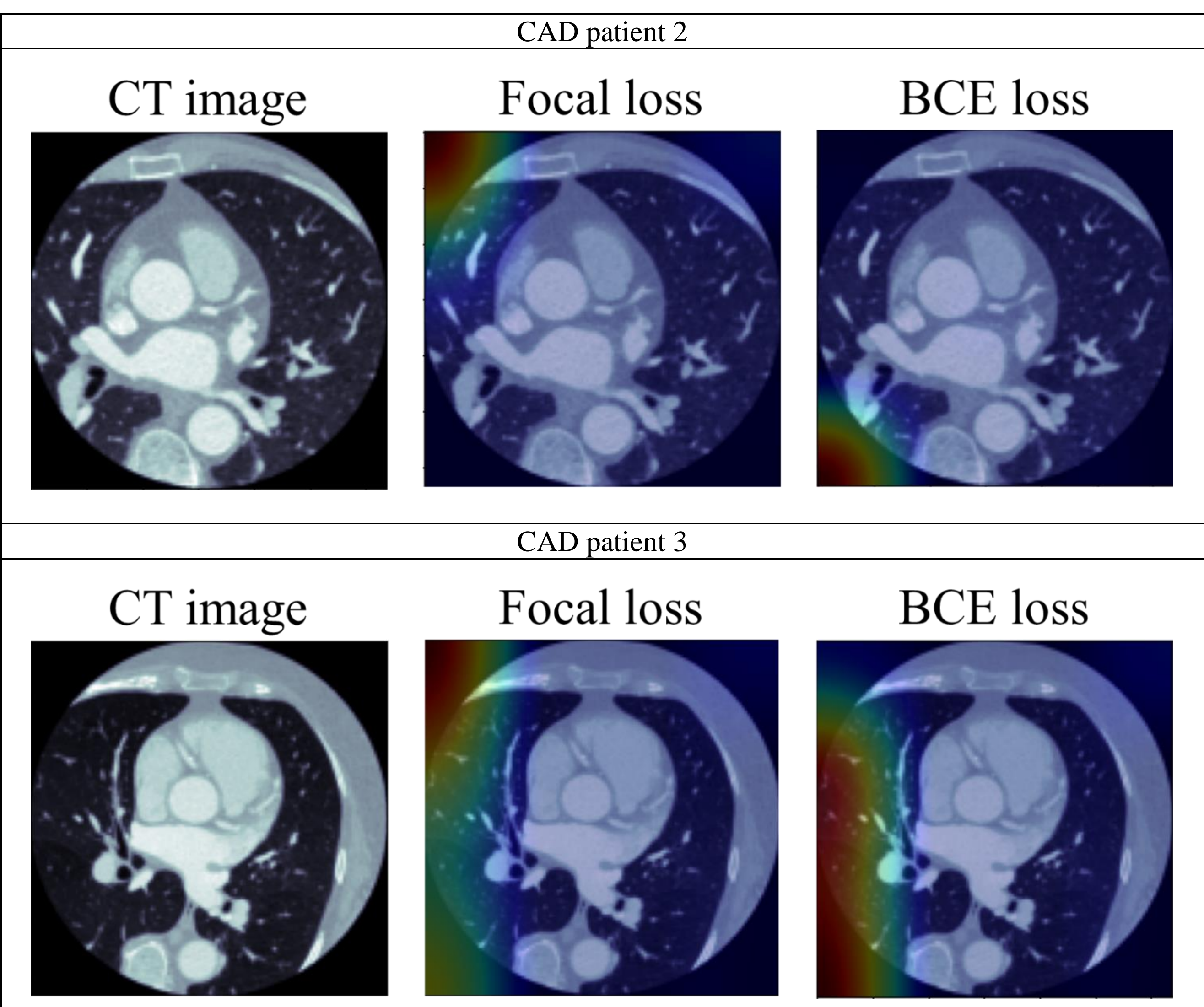

Figure 4: The Grad-GAM generated by the final 2D Resnet-50 model from last convolution layer on a CAD patient 2 and a CAD patient 3 with incorrect prediction.

(2) 3D CAD classification by the final 3D Resnet-50 model

Regarding the correct prediction, the 3D model with FL focused on the heart, including coronary arteries for the normal subject. On the other hand, the 3Dmodel with BCE focused on part of the heart only. For CAD patient 1, the 3D model with FL also focused on the heart, including coronary arteries, while the 3D model with BCE focused on the right lung and did not focus on the heart.

Regarding the incorrect prediction, the 3D model with FL still focused on the heart, including coronary arteries for both patients, while the 3D model with BCE focused on the bottom region of the CT image and did not focus on the heart.

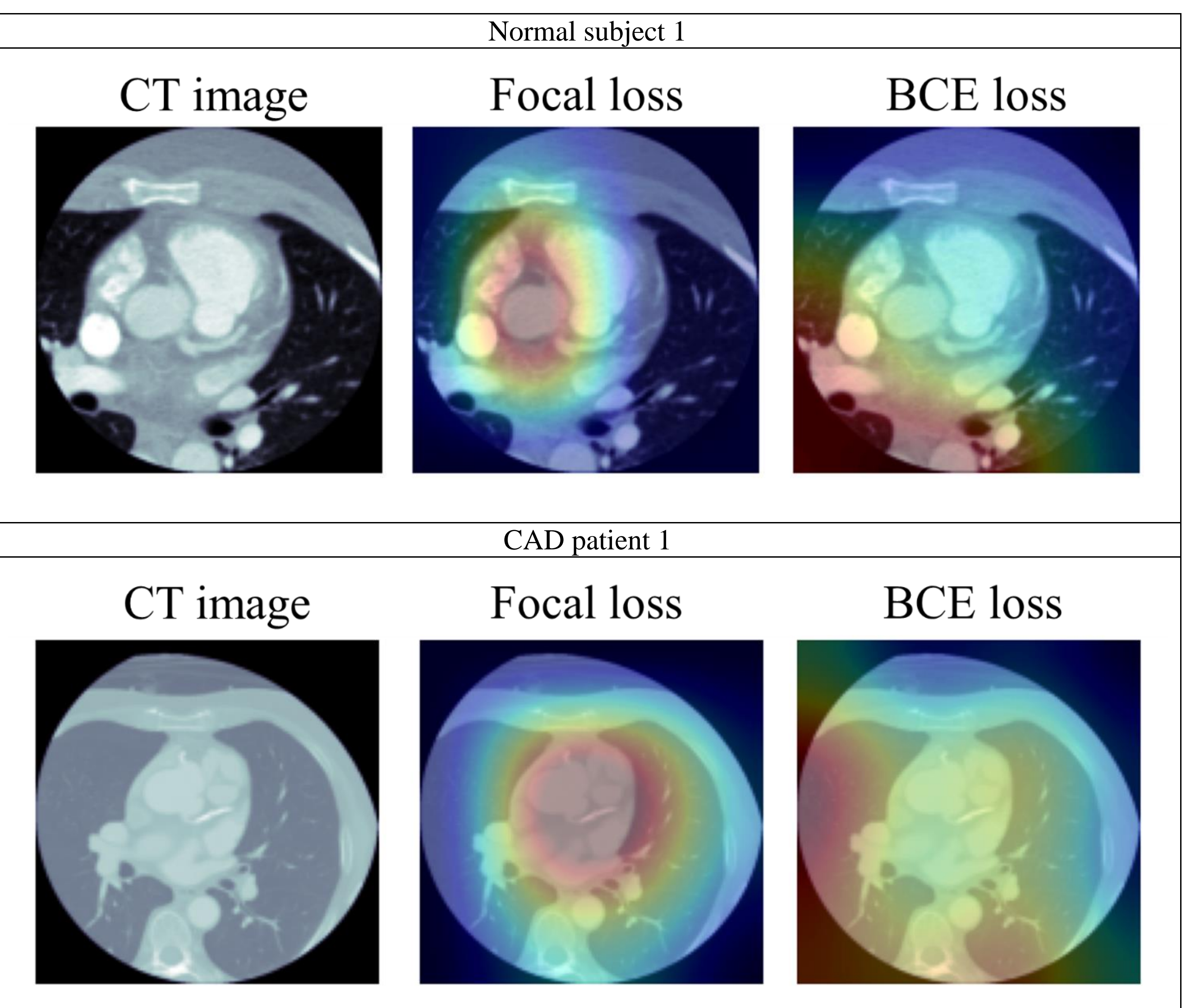

Figure 5: The Grad-GAM generated by the final 3D Resnet-50 model from last convolution layer on a normal subject 1 and a CAD patient 1 with a correct prediction.

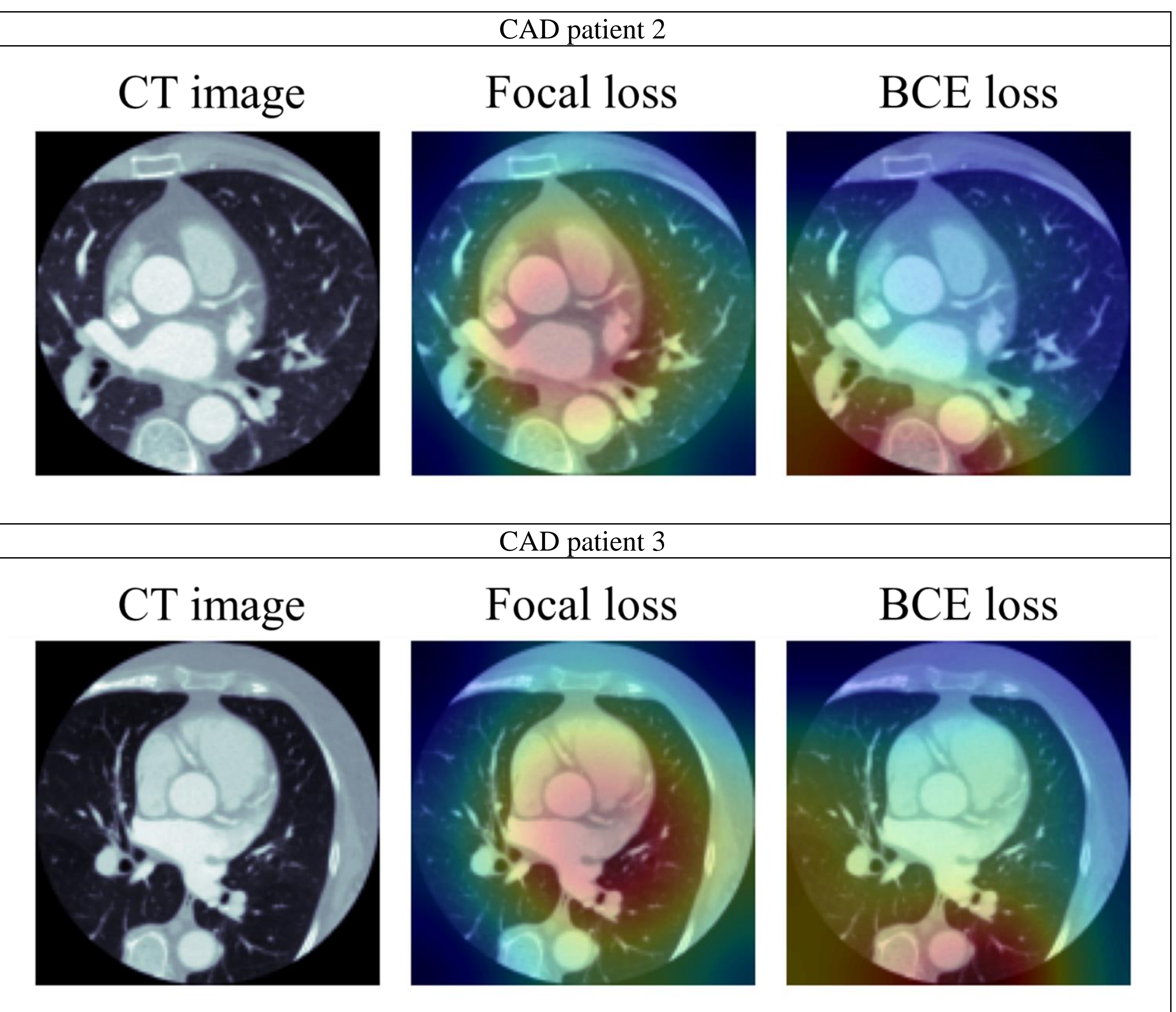

Figure 6: The Grad-GAM generated by the final 3D Resnet-50 model from last convolution layer on a CAD patient 2 and a CAD patient 3 with incorrect prediction.

(3) 2D two-class semantic segmentation by 2D modified U-Net model

Figures 7 and 8 shows the aorta and coronary arteries segmentation and its explainability. It confirms that the deep learning model focused on the aorta and coronary arteries for the segmentation task. Further, the modified U-Net showed a good segmentation performance (DSC = 91.2%). We observed that a 2D deep learning based semantic (two-class) segmentation provides a better solution for accurate classification and explainability with exact abnormality localisation.

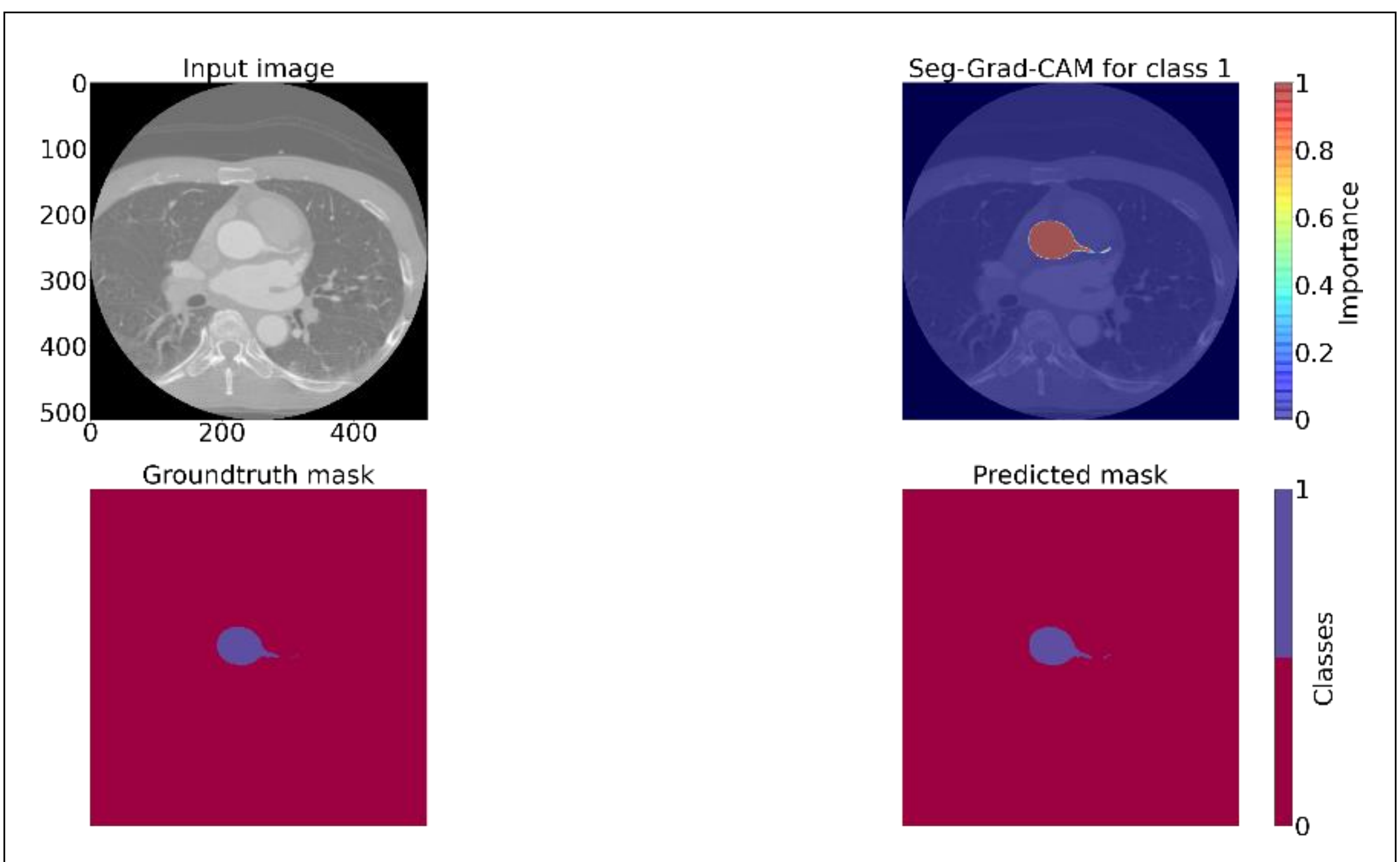

Figure 7: An example of selected CTCA slice and its ground-truth mask, Seg-Grad-CAM for aorta and coronary arteries and predicted mask. Part of the image is reproduced with permission from [26].

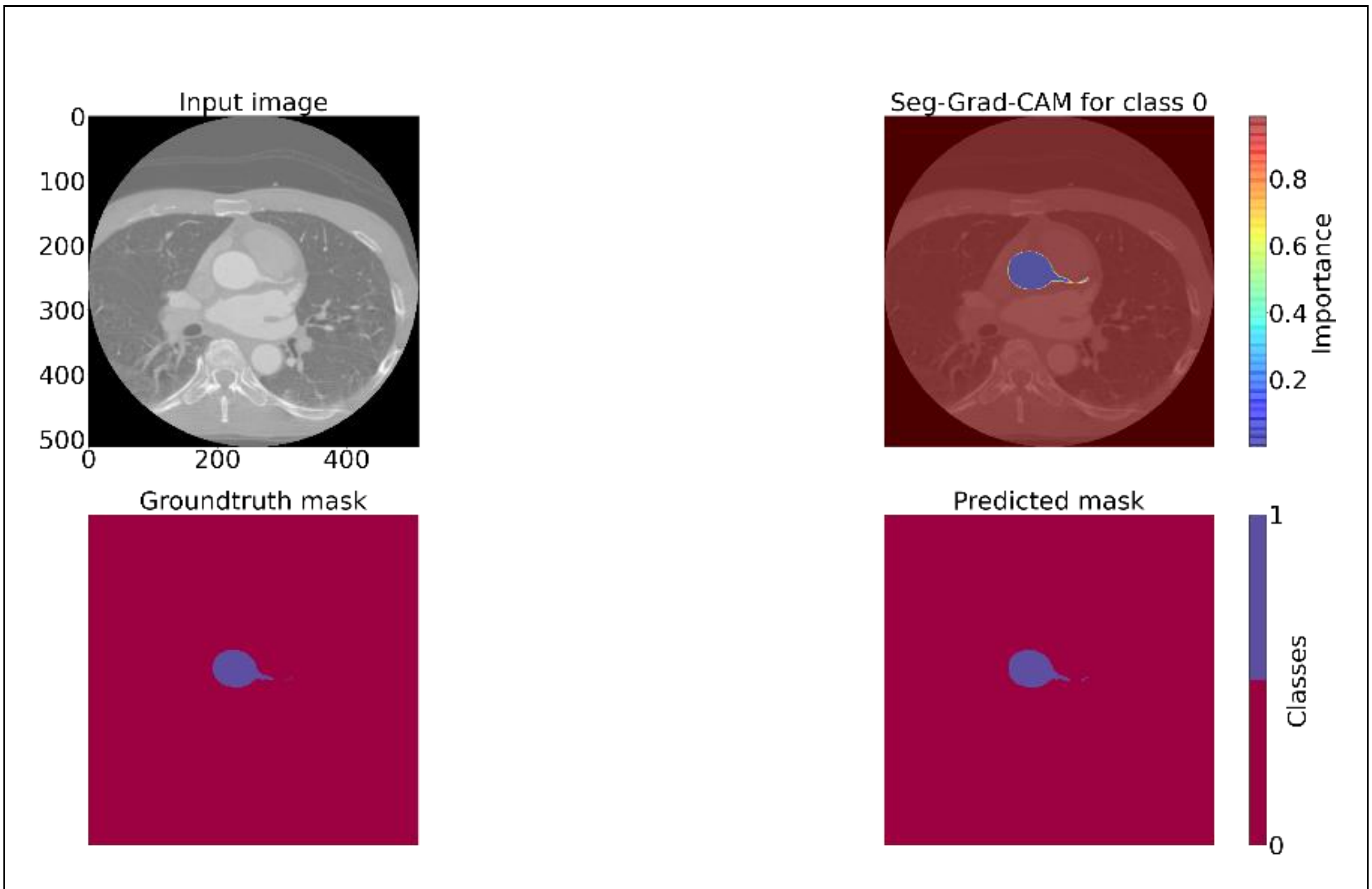

Figure 8: An example of selected CTCA slice and its ground-truth mask, Seg-Grad-CAM for non-aorta and non-coronary arteries and predicted mask. Part of the image is reproduced with permission from [26].

*Comparative analysis of SOTA methods*

The recall, precision, F1-score and accuracy of the 3D DenseNet-121, 3D EfficientNetB0, and the proposed 3D Resnet-50 models for test set are shown in Table 9. With regard to recall, precision, F1-score, and accuracy measurements on the test set, our proposed model outperformed the 3D DenseNet-121 and 3D EfficientNetB0 models by 21.43% on test accuracy. Our approach with FL achieved 0.43, 1.00, and 0.60 for recall, precision, and F1-score, respectively, while 3D DenseNet-121 and 3D EfficientNetB0 achieved 1.00, 0.50, and 0.67 for recall, precision, and F1-score, respectively.

Table 9: Comparison with SOTA models for test set.

| Final classification model | Recall | Precision | F1-score | Accuracy |
|---|---|---|---|---|
| 3D DenseNet-121 (FL) | 1.00 | 0.50 | 0.67 | 50.00% |
| 3D EfficientNetB0 (FL) | 1.00 | 0.50 | 0.67 | 50.00% |
| Proposed 3D Resnet-50 (FL) | 0.43 | 1.00 | 0.60 | 71.43% |

**Discussion**

A 3D deep learning-based classifier has been proposed to classify CAD patients from CTCA images. It adopts a 3D Resnet-50 model with FL to classify normal subjects and patients with CAD. Its classification accuracy is better than SOTA models by 21.43%. Furthermore, our proposed model provides better explainability, aligns with clinicians' expectations as well as their diagnostic criteria. It also has excellent precision and fair recall, which implies that it does not misclassify normal subjects as CAD patients and is good at identifying normal subjects.

In this study, we observe that overfitting of a 2D Resnet-50 model can lead to poor classification performance. The overfitting is due to memorisation of inaccurate labels. The classification performance could be improved by introducing an automatic inaccurate label correction method [27]. It should be noted that our previous study [28] produced better classification accuracy than the current study as it employed unbalanced datasets for training, validation and test. Therefore, the previous results were biased towards CAD cases. Our current study employed balanced datasets for all sets, and therefore produced unbiased results for both normal and CAD cases.

It should be noted that the 2D model was employed to classify the CTCA image per slice only. We do not intend to use this model to perform patient-level classification, and therefore techniques such as maximum voting or average voting are not used. Rather, the performance of this model can be served as a reference when incorrect labels were utilised. Interestingly, its performance is better than a purely random guess.

The five-fold cross-validation results showed that the validation accuracy is higher than the training accuracy for all models. It confirms that the dropout layer employed in our 3D model could provide better generalisation on the validation set.

Explainability has been reported in this study. In terms of abnormality localisation, our proposed 3D model with FL produced a focused heat map (i.e., focused on the heart, including coronary arteries) on the last convolution layer, while the same model with BCE did not produce a focused heat map. Furthermore, our study confirms that a model having good classification accuracy does not necessarily have good explainability; the choice of the loss function is crucial not only for classification but also for explainability. This also reveals that our 3D model with FL has the ability to capture small plaques and stenosis in CTCA images. Most current methods require more complex procedures, such as (1) localising the coronary arteries by centerline extraction and MPR, and (2) detecting stenosis and plaque. On the other hand, our proposed method, when compared with the current methods, provides a direct and simpler approach for CAD classification.

Recently, Ghassemi et al. [29] have discussed the inaccuracy of the heat map. Given that the annotation is at patient-level, the generated heat map does not highlight the abnormality exactly, limiting interpretability for clinicians. Based on our observations in this study, we deduce that the accuracy/usefulness of the explainability depends on the level of annotation (i.e., annotation at patient level vs. annotation at pixel/voxel level). An alternative way to increase the annotation level is to incorporate prior information into the deep learning model (i.e., informed cue [30]). In the context of CAD classification, the cue could be a mask containing the aorta and coronary arteries. The cue could help a deep learning model focus on important regions relevant to clinicians' diagnostic criteria. The associated heat map could provide better explainability, thereby meeting clinicians' expectations.

We demonstrate the reasoning above by treating 3D CAD classification as a 2D two-class (aorta and coronary arteries vs non-aorta and non-coronary arteries) semantic segmentation. It confirms that the deep learning model focused on the aorta and coronary arteries for the segmentation task. We conclude that a 2D deep learning based semantic (two-class) segmentation provides a better solution for accurate classification and exact explainability with exact abnormality localisation.

In practice, we suggest that the design of deep learning methods and their explainability should meet the clinicians' requirements for accuracy and explainability. For example, CAD severity classification can be performed first, and coronary artery segmentation can be performed subsequently. In this approach, the clinicians will be informed not only of the CAD severity and the exact location of the abnormality but also of the exact explainability. This could also improve the trust and transparency of deep learning models and make them easier to be deployed in the clinical setting.

Another observation of our study is that the proposed 3D model has shorter training time when compared with the 2D model. It requires only ten epochs to learn the image features and demonstrates good classification performance as well as explainability.

Our research has practical implications for healthcare services. This study provides a fast, efficient, objective, and accurate automated detection and classification method of coronary artery disease in an Accident and Emergency department using a computer-based deep learning algorithm to improve CAD diagnosis. It will not only enable fast triage of patients using this technology on local hospital servers, speeding up diagnosis and patient management, but also reduce the healthcare cost.

The study has several limitations. First, the study is retrospective, the selection of patients might be biased. Second, the slice thickness might limit the visibility of very small coronary arteries on CTCA images. Lastly, the relatively small dataset might limit the generalisation of the proposed deep learning model.

In future work, we hope to incorporate the mask of the aorta and coronary arteries into 3D CAD classification and further improve classification accuracy and explainability. The 2D 2-class semantic segmentation can be extended to 3D multi-class (i.e., aorta and coronary arteries, stable plaques and unstable plaques) segmentation; a data-centric strategy [31] can be deployed for more accurate CAD classification and explainability with exact abnormality localisation.

## Conclusion

A 3D deep learning-based CAD classifier with good accuracy and explainability has been proposed. Our method outperformed the SOTA methods in terms of classification accuracy. Furthermore, we demonstrated that CAD semantic segmentation could provide exact explainability that aligns with clinicians' diagnostic criteria. We contribute to the human-centric healthcare AI research by developing and delivering AI-enabled CAD classification tools with explainability. Our research has the potential impact of improving the healthcare service and reducing its cost.

## Acknowledgement

The authors acknowledge Mr Xiaoxia Yang for his assistance with data processing for aorta and coronary artery segmentation. JJ was supported by Wellcome Trust Clinical Research Career Development Fellowship 209553/Z/17/Z and the NIHR Biomedical Research Centre at University College London. This research was funded in whole or in part by the Wellcome Trust [209553/Z/17/Z]. For the purpose of open access, the author has applied a CC-BY public copyright licence to any author accepted manuscript version arising from this submission.

## Declaration of Interest

JJ declares fees from Boehringer Ingelheim, F. Hoffmann-La Roche, GlaxoSmithKline, NHSX, Takeda, Wellcome Trust, Microsoft Research unrelated to the submitted work and UK patent application numbers 2113765.8 and GB2211487.0.